\documentclass[journal = acs-nl, manuscript = article]{achemso}

\usepackage{graphicx}
\usepackage{hyperref}
\usepackage{natbib}
\usepackage{booktabs}
\usepackage{amsmath}
\usepackage{amssymb}
\usepackage{color}
\usepackage{xcolor}
	\title{{\color{black}Mass inversion at the Lifshitz transition in monolayer graphene by diffusive, high-density, on-chip, doping}}
	
	\author{Ayse Melis Aygar}
	\affiliation{Dept. of Electrical and Computer Engineering, McGill University, Qu\'ebec, Montr\'eal, H3A-0E9, Canada}
	\author{Oliver Durnan}
	\affiliation{Dept. of Electrical Engineering, Columbia University, New York, NY, 10027, USA}
\author{{\color{black}Bahar Molavi}}
	\affiliation{Dept. of Electrical and Computer Engineering, McGill University, Qu\'ebec, Montr\'eal, H3A-0E9, Canada}	
\author{{\color{black}Sam N. R. Bovey}}
	\affiliation{Dept. of Electrical and Computer Engineering, McGill University, Qu\'ebec, Montr\'eal, H3A-0E9, Canada}
	\author{Alexander Gr\"uneis}
	\affiliation{Institut f\"ur Festk\"orperelektronik, Technische Universit\"at Wien, Vienna, 1040, Austria}
	\author{Thomas Szkopek}
	\affiliation{Dept. of Electrical and Computer Engineering, McGill University, Qu\'ebec, Montr\'eal, H3A-0E9, Canada}
\email{thomas.szkopek@mcgill.ca}
	
	\date{\today}
	
\begin{document}

	\begin{abstract}
	Experimental setups for charge transport measurements are typically not compatible with the ultra-high vacuum conditions for chemical doping, limiting the charge carrier density that can be investigated by transport methods. Field-effect methods, including dielectric gating and ionic liquid gating, achieve {\color{black}too low a carrier density to induce electronic phase transitions}. To bridge this gap, we {\color{black}developed} an integrated flip-chip method to dope graphene by alkali vapour in the diffusive regime, suitable for charge transport measurements at ultra-high charge carrier density. We introduce a cesium droplet into a sealed cavity filled with inert gas to dope a monolayer graphene sample by the process of cesium atom diffusion, adsorption and ionization at the graphene surface, with doping {\color{black}beyond an electron density of $4.7\times10^{14}~\mathrm{cm}^{-2}$} monitored by operando Hall measurement. The sealed assembly is stable against oxidation, enabling measurement of charge transport versus temperature and magnetic field. {\color{black}Cyclotron mass inversion is observed via the Hall effect, indicative of the change of Fermi surface geometry associated with the Liftshitz transition at the hyperbolic $M$ point of monolayer graphene. The transparent quartz substrate also functions as an optical window, enabling non-resonant Raman scattering.} Our findings show that chemical doping, hitherto restricted to ultra-high vacuum, can be applied in a diffusive regime at ambient pressure in an inert gas environment {\color{black}and thus enable charge transport studies in standard cryogenic environments.}

	\end{abstract}
		
	\section{Introduction}
	
\begin{figure*}[!] 
		\begin{center}
			\includegraphics[width=0.8\textwidth]{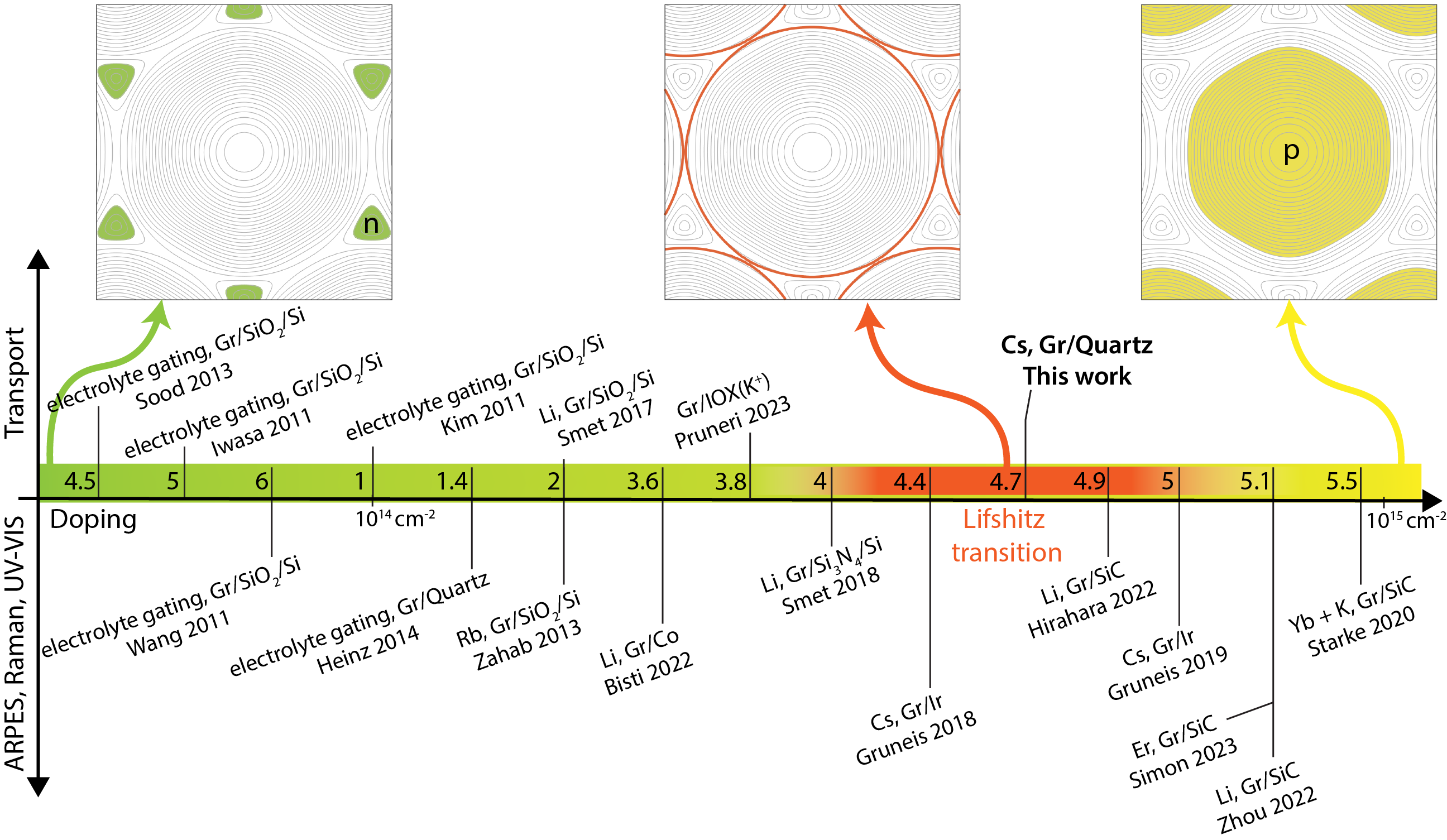}
			\caption{\textbf{Monolayer graphene doping and the Lifshitz transition} A schematic of the Fermi surface contours in $\pi\ast$ conduction band of monolayer graphene vs. doping. Electron and hole pockets are shown in green and yellow, respectively. The Lifshitz transition is estimated to occur at $n=3.7-5.1\times10^{14}~\mathrm{cm}^{-2}$ \cite{rosenzweig2020overdoping,zaarour2023flat}. The charge carrier densities achieved in ARPES, Raman and UV-VIS  (below) and charge transport studies (above) are summarized.}
			\label{fig1}
		\end{center}	
	\end{figure*}

{\color{black}The electron transport properties of nearly charge neutral graphene have been the subject of intense investigation. Much less is know about the electron transport properties of heavily electron doped graphene, where the massive, hyperbolic dispersion at the $M$ saddle points in the Brillouin zone lead to a van Hove singularity (vHS) in the density of states \cite{mcchesney2010extended, kravets2010spectroscopic, mak2011seeing, mak2014tuning}. Angle resolved photoemission spectroscopy (ARPES) studies have mapped the Fermi surface of heavily doped monolayer graphene in the vicinity of the $M$ point \cite{mcchesney2010extended,hell2018resonance,Ehlen:2020,rosenzweig2020overdoping,zaarour2023flat}. The role of many-body interactions in flattening and thereby extending the vHS is a matter of active debate\cite{bao2022coexistence,ichinokura2022van,jugovac2022clarifying}. Indeed, it can be difficult to probe many-body effects by ARPES due to the flattness of the band, further motivating investigation by complementary methods such as charge transport. The emergence of superconductivity in doped monolayer graphene prompted by strong electron-phonon coupling at the extended vHS has been the subject of theoretical inquiry \cite{Profeta:2012,Nandkishore:2012} and experimental evidence of electron-phonon coupling enhancement has been reported \cite{ludbrook2015evidence}. Charge transport is one of several physical properties that are sensitive to changes in the Fermi surface geometry at a vHS \cite{kaganov1979electron}, known as a Lifshitz transition (Fig. \ref{fig1}), yet no transport measurements in heavily doped graphene in the vicinity of the vHS have been reported to date.}

Alkaline metal doping (Li, K, Ca, Cs) and rare earth doping (Er, Yb) of graphene in an ultra-high vacuum (UHV) environment has enabled doping in excess of $n=4\times10^{14}~\mathrm{cm}^{-2}$ for surface science studies of graphene \cite{mcchesney2010extended,hell2018resonance,Ehlen:2020,rosenzweig2020overdoping,ichinokura2022van, bao2022coexistence,zaarour2023flat}. However, UHV environment requirements are rarely fulfilled in charge transport measurements, particularly in high magnetic field experiments. Li doping of epitaxial graphene on SiC in an integrated deposition and cryogenic charge transport measurement system was used to achieve electron densities as high as $2.2\times10^{14}~\mathrm{cm}^{-2}$ \cite{khademi2019weak}. Ionic liquid gating has enabled electron doping in the range of $0.5-2\times10^{14}~\mathrm{cm}^{-2}$ in electrically contacted graphene on Si/SiO$_2$ substrates \cite{efetov2010controlling,ye2011accessing,Chen:2011}. {\color{black}Ion exchange glass has achieved doping up to $3.8\times10^{14}~\mathrm{cm}^{-2}$ with potassium\cite{graham2023highly}.} Lithium polymer electrolytic gating has been used to achieve electron doping as high as $2\times10^{14}~\mathrm{cm}^{-2}$ in electrically contacted bilayer graphene \cite{Kuhne:2017} {\color{black}and can be integrated with magnetotrasport measurements.} Transmission electron microscopy reveals domains with superdense Li ordering in suspended graphene bilayers corresponding to $n=4\times10^{14}~\mathrm{cm}^{-2}$ per graphene sheet \cite{kuhne2018reversible}, {\color{black}below the monolayer vHS threshold. There is thus a need to develop new methods that combine the advantages of chemical doping, high electron density and selection of chemical species, that can be easily integrated with existing charge transport experimental environments.}
	
	We present here an integrated flip-chip method to dope graphene by alkali vapour in the diffusive regime, suitable for charge transport measurements at ultra-high charge carrier density. The method enables operando electronic characterization and allows the hermetically sealed device to be analyzed without the requirement of UHV or inert gas environmental conditions. We demonstrate our method by doping electrically contacted large-area graphene to ultra-high charge carrier density, {\color{black} reaching $n=4.7 \times 10^{14}~\mathrm{cm}^{-2}$ as confirmed by the inversion of effective mass in high-field Hall measurements.} This flip-chip method is versatile: it can be applied to different host material and dopant systems, {\color{black}and is not limited to monolayer or few-layer host materials.}

The method is compatible with operando charge transport characterization during alkali doping. As shown in Fig. \ref{fig2}a), flip-chip encapsulation of an alkaline metal vapour source in an inert gas environment electron dopes electrically contacted graphene. The impermeability of the encapsulation to external oxidizing agents permits sample manipulation in an ambient atmospheric environment, enabling the study of heavily doped graphene by different experimental techniques. We report here operando measurements of the graphene resistivity tensor elements $\rho_{xx}$ and $\rho_{xy}$ in an AC magnetic field during the doping process, as well as the measurement of Hall effect, weak-localization and magnetoresistance at cryogenic temperature $T= 1.3$~K at magnetic fields up to $B=7$~T. {\color{black}Cryogenic Hall measurements reveal inversion of the Hall coefficient, corresponding to the inversion of the cyclotron effective mass $m_* =  \hbar^2 (2\pi)^{-1} \partial A_k / \partial E_F$, where $A_k$ is the area subtended by the Fermi-surface \cite{ashcroft1976solid}, as expected at the Lifshitz transition at the hyperbolic $M$ point of heavily doped graphene (Fig. \ref{fig1}). Non-resonant Raman measurement of heavily doped graphene through the optically transparent quartz substrate reveals a $G$ peak Raman shift in agreement with previous reports of heavily doped graphene \cite{hell2018resonance}.}

	\section{Experimental Methods}

	We used graphene grown by chemical vapour deposition (CVD) on poly-crystalline copper foil in a cold wall CVD reactor (Aixtron BM) at Graphenea, and transferred onto a 100~mm diameter quartz (001) substrate following sacrificial etch of the copper growth substrate. {\color{black}The graphene crystal domain size is estimated to be 1.5~$\mu$m by Raman spectroscopy (see Supplementary Information).} Quartz is the substrate of choice due to its chemical inertness and optical transparency. We prepare $50~\mathrm{\mu m} \times 500~\mathrm{\mu m}$ graphene Hall-bar devices using photo-lithographic methods, with Ti/Au (5 nm /80 nm) Ohmic contacts. {\color{black}Sonication in an acetone bath, followed by $>24$ hour immersion in 1,2 dichloroethane is required to minimize organic residue contamination that may inhibit charge transfer.} The six-contact Hall-bar geometry enables the simultaneous measurement of four-probe longitudinal resistance $R_{xx}=V_x/I_x$ and transverse resistance $R_{xy}=V_y/I_x$.
		
\begin{figure}[!]
		\begin{center}
			\includegraphics[width=0.7\columnwidth]{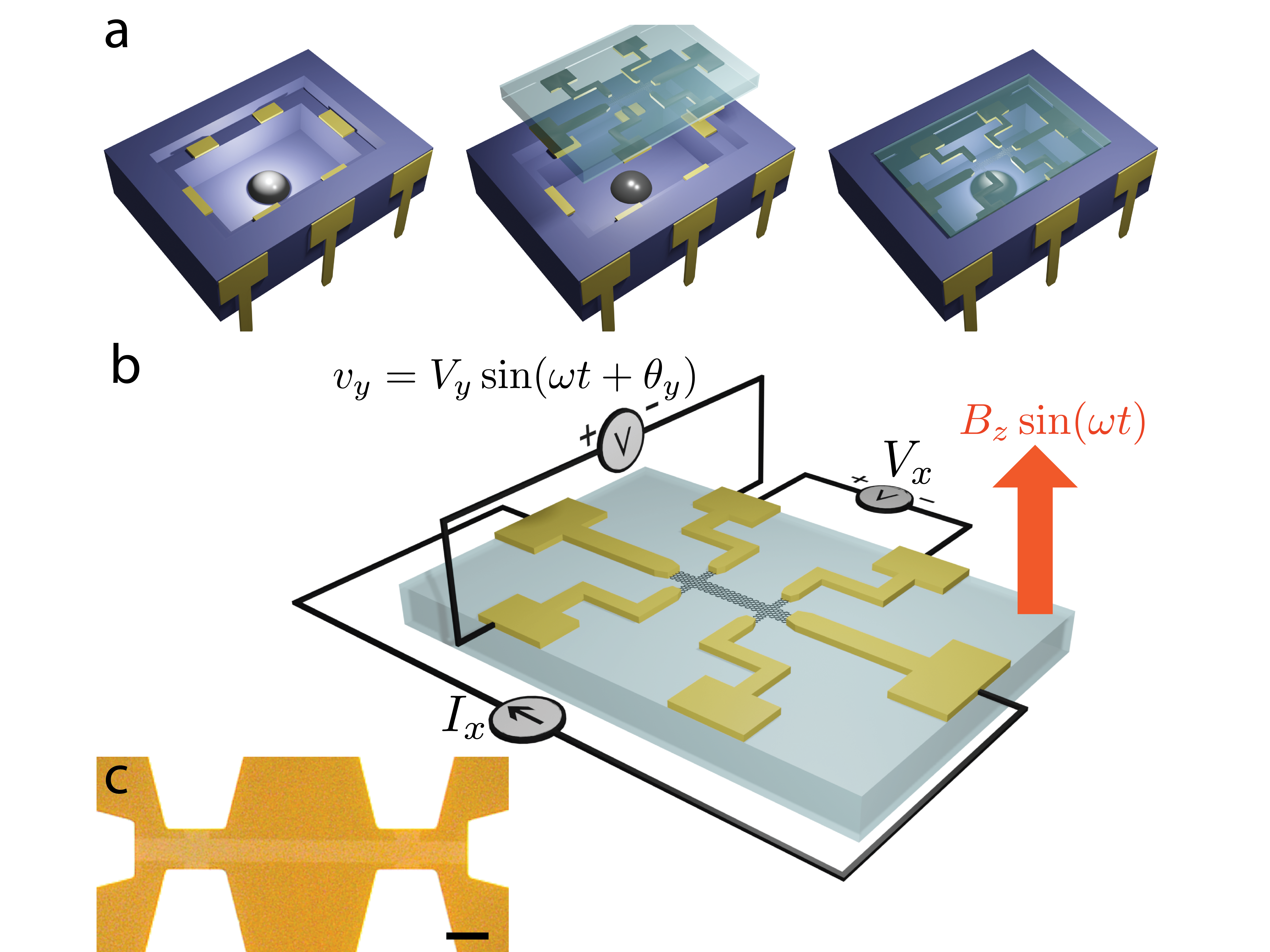}
			\caption{\textbf{Flip-chip alkali doping method} (a) Schematic of the flip-chip assembly method. A drop of Cs is placed in the chip carrier. A graphene Hall bar on quartz is brought face down to enclose the Cs in the cavity and expose graphene to Cs vapour. (b) A schematic of the AC Hall measurement configuration for the Hall bar on quartz. (c) Optical microscope image of a graphene Hall bar on quartz. The scale bar corresponds to 100~$\mathrm{\mu m}$.}
			\label{fig2}
		\end{center}
	\end{figure}

	The flip-chip method uses a ceramic chip carrier with a cavity and gold electrodes atop a recessed mesa (Fig. \ref{fig2}a). A drop of Cs ($\approx 10~\mu$L in volume) is drop-cast by glass pipette into the cavity inside an {\color{black}inert gas} glove box (H$_2$O $\le$ 0.1 ppm and O$_2$ $\le$ 1 ppm). {\color{black} Prior to all doping experiments, we thermally annealed the Hall bar samples and chip carriers in the inert glove box environment for $>2$ hours at 120$^\circ$C to remove adsorbed water. Experiments conducted in an Ar environment enabled higher doping density to be achieved than in an N$_2$ environment.} The quartz substrate with the graphene Hall bar is flipped onto the chip carrier. Pure In spheres (0.5~mm diameter) are used for reliable electrical contact between the electrodes of the chip carrier and the quartz substrate. The graphene Hall bar faces the Cs source directly and is exposed to Cs vapour within the cavity. A compact resistive heater under the chip carrier is used to increase temperature and thus Cs vapour pressure. Interestingly, these experiments are carried out in a 1 atm inert gas environment, which limits the mean free path of the Cs atoms, corresponding to a diffusive vapour transport regime as opposed to the ballistic regime of alkali doping in UHV conditions. Despite this significant difference in the vapour transport regime, we observe that similar charge carrier densities can be achieved.
	
	An electromagnet was used for operando Hall measurement during doping in a glove box. We used an AC-DC Hall measurement configuration (Fig. \ref{fig2}b) with an AC magnetic field, $B_z \sin(\omega t)$ of amplitude $B_z$ = 20-25~mT and frequency $f = 3$~Hz, and a DC sample bias current, $I_x= 200~\mu\mathrm{A}$. A similar method has been employed for Hall measurement of organic field effect transistors \cite{chen2016high}. The Hall voltage is expected to be $V_{y} \sin(\omega t+\theta _y)$ where $\theta _y$ is 0 or $\pi$ according to charge carrier type. In all experiments, the direction of the bias current is selected such that Hall voltage phase $\theta_y=0$ ($\theta_y=\pi$) corresponds to electron (hole) doped graphene. The background inductive voltage was subtracted to obtain sample response.

	Following Cs doping, we sealed the flip-chip cavity with a UV activated epoxy (Solarez) followed by potting epoxy compound (Loctite Stycast). {\color{black}The adhesives were degassed of oxidizing vapours in an inert gas environment prior to use.} The sealed device was removed from the glove box for high magnetic field transport measurements and Raman spectroscopy. A representative optical image of a graphene Hall bar on quartz substrate is shown in Fig. \ref{fig2}c. High magnetic field ($B$ = 7~T) and low-temperature ($T$ = 1.3~K) measurement of longitudinal and transverse resistance was performed in a closed-cycle cryostat with a superconducting solenoid. Raman spectroscopy was performed with a Renishaw InVia confocal Raman microscope, 50-X magnification, $\lambda = 785~$nm, $P = 90$~mW.

	\section{Results}
	CVD grown graphene transferred onto quartz is initially p-type doped {\color{black}($p \approx 0.9-1.6\times10^{12}~\mathrm{cm}^2$, $\mu = 2200-3100~\mathrm{cm}^2\mathrm{V}^{-1}\mathrm{s}^{-1}$)}, as measured by Hall effect in our samples prior to doping. During exposure to Cs vapour, ionization of adsorbed Cs atoms leads to valence electron donation to the graphene. {\color{black}The resistances $R_{xx}$ and $R_{xy}$ are shown in Fig. \ref{highdoping}a,b during a representative Cs vapour exposure experiment. Doping is sufficiently rapid that conduction is already n-type at the beginning of the measurement. A slower doping rate in a less thoroughly cleaned graphene sample allows observation of passage through the charge neutrality point (see Supplementary Information). The Cs doping process is partially reversible, achieved by heating a doped graphene sample in Cs free chipe carrier (see Supplementary Information). The Hall coefficient $R_H = -\partial{R_{xy}}/\partial{B_{z}}$ can be used to directly infer charge carrier density in the case of conduction by a single charge carrier type, with $n = -1/eR_H$ for n-type graphene. The quantity $-1/eR_H$ and the Hall mobility $\mu = R_{xy}/(B_{z}\rho_{xx})$, where $\rho_{xx}$ is the graphene resistivity determined from $R_{xx}$ and Hall bar geometry, versus time are shown in Fig. \ref{highdoping}c,d. The chip carrier was heated to elevate Cs temperature, and thus Cs vapour pressure, doping rate, and doping density. Note that the Hall coefficient $R_H$ becomes sufficiently small upon doping that the noise limit of the operando Hall measurement is reached, and the inferred electron density $-1/eR_H = 7\times10^{14}~\mathrm{cm}^{-2}$ exceeds the stoichiometric limit of $n=4.77\times10^{14}~\mathrm{cm}^{-2}$ corresponding to CsC$_8$ observed in UHV Cs-doping experiments\cite{Ehlen:2020}. We attribute this observation to the onset of partially compensated conduction of charge carriers with positive and negative effective masses at the vHS, requiring investigation at higher magnetic field.

	\begin{figure}[!]
		\includegraphics[width=0.6\columnwidth]{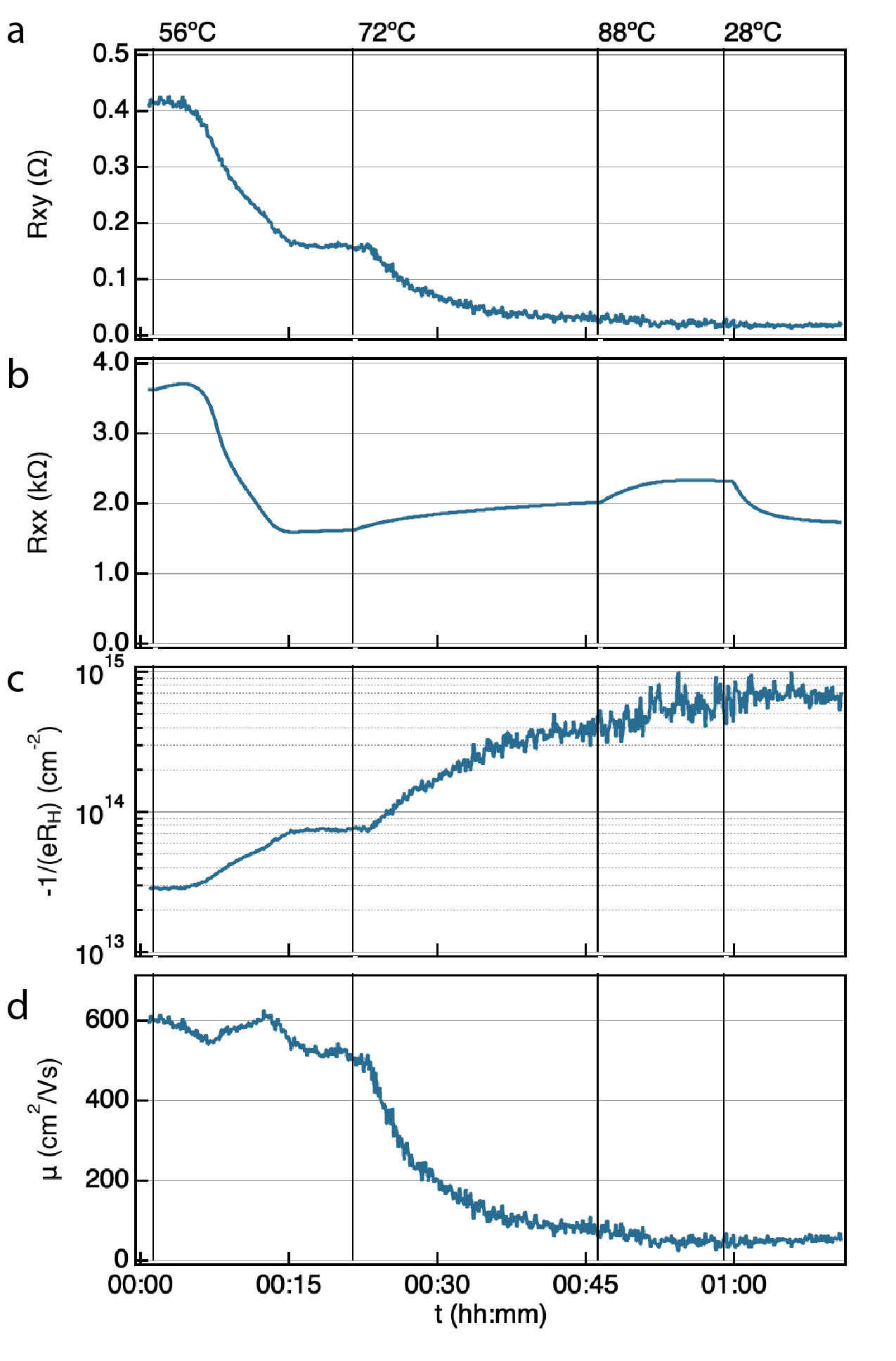}
		\caption{\textbf{Operando Hall measurement of graphene doping}.(a) The Hall resistance $R_{xy}$ and (b) the longitudinal resistance $R_{xx}$ versus time $t$ during Cs vapour exposure. The rate of Cs doping is increased by heating the chip carrier, with sample temperature set-points for different time intervals indicated. (c) The inverse Hall coefficient, $1/eR_H$ and (d) the Hall mobility $\mu$ inferred from $R_{xy}$ and $R_{xx}$ versus time $t$ reveal in the increase in electron density $n$ and reduction in mobility. The inverse Hall coefficient reaches $1/eR_H = 7\times10^{14}\mathrm{cm}^{-2}$, beyond the expected limit of electron density in graphene, implying the onset of compensated electron and hole conduction.}
		\label{highdoping}
	\end{figure}

Magnetotransport experiments were conducted at high field ($B = 7$~T) and cryogenic temperature ($T = 1.3$~K). The Hall resistance $R_{xy}$ versus $B$ is shown in Fig. \ref{fridge}a for five samples (identified 1-5), with an inset of the low $R_{xy}$ regime. All samples show a reduction in $R_{xx}$ as temperature decreases (Fig. \ref{fridge}b), as well as a weak localization peak and small magnetoresistance in $R_{xx}$ versus $B$ (Fig. \ref{fridge}c). Table \ref{tab:table1} presents a summary of the inferred charge carrier density $n$, Hall mobility $\mu$ and other properties of samples 1-5. Samples 1-3 exhibit n-type behaviour with charge carrier density up to $n=4.77\times10^{14}~\mathrm{cm}^{-2}$. Samples 4 and 5 exhibit a reciprocal Hall coefficient $1/R_H$ too large to be attributed to a single charge carrier type, with sample 5 exhibiting an \textit{inversion} in the sign of the Hall coefficient corresponding to a change from n-type to p-type conduction. 

To gain further insight into the electronic properties of these heavily doped samples, a third-nearest neighbour tight-binding (3NNTB) calculation of electronic band structure using hopping and overlap parameters determined from ARPES\cite{gruneis2008tight} was performed. 
From the 3NNTB electronic structure and experimental electron density $n$, the Fermi level $E_F$ was inferred, and Fermi surfaces identified in the graphene $\pi*$ band (Fig. \ref{fridge}d,e). The compensated conduction of samples 4 and 5 correspond to a Fermi surface below and above, respectively, the Lifshitz transition where $E_F = E_M = 1.93$~eV. The Hall coefficient in the vicinity of the Lifshitz transition can be understood with a simple two-band model \cite{ashcroft1976solid} in the low-field limit,
\begin{equation}
R_H = \frac{\sigma^2_n R_n + \sigma^2_p R_p}{(\sigma_n+\sigma_p)^2}
\label{twoband}
\end{equation}
where $R_j$ and $\sigma_j$ are the Hall coefficients and conductivities of two $j=n,p$ bands. Electrons at $E_F = 1.93 - \delta$~eV for small $\delta$ are characterized by a Hall coefficient $R_n = -1/(2 e A_K) = -4.74\times 10^{14}~\mathrm{cm}^{-2}$, where $A_K$ is the area around the $K$ and $K'$ points enclosed by the Fermi surface in the first Brillouin zone. The factor of 2 accounts for spin degeneracy, and the numerical value of $R_n$ is inferred from 3NNTB calculation. Electrons at $E_F = 1.93 + \delta$~eV for small $\delta$ are characterized by a Hall coefficient $R_p = 1/(2 e A_\Gamma) = 3.34\times 10^{15}~\mathrm{cm}^{-2}$, where $A_\Gamma$ is the area enclosing $\Gamma$ with opposite cyclotron motion to that of $R_n$. Equivalently, the effective cyclotron mass $m_*/m_0 =  \hbar^2 (2\pi m_0)^{-1} \partial A_k / \partial E_F$ reverses sign as $E_F$ crosses the vHS beyond the equi-energy contours connecting the $M$-points, and $A_k$ undergoes a discontinuous change from an n-type $A_K$ to a p-type $A_\Gamma$. With the two-band model of Eq. \ref{twoband}, we find that $\sigma_p/\sigma_n = 0.13$ and $3.07$ for samples 4 and 5, respectively, corresponding to net n-type and p-type behaviour in the vicinity of the vHS. As electron doping increased, conduction in the electron pockets around $K$ and $K'$ is closed and conduction in the hole pocket around $\Gamma$ is opened.

	\begin{table*}[t]
		\begin{center}
			\caption{\textbf{Physical properties of highly doped graphene}. Various experimentally measured and model parameters for samples 1-5, including inverse Hall coefficient $(eR_H)^{-1}$, inferred doping density $n$, the Hall mobility $\mu$, the estimated CsC$_\mathrm{x}$ coverage estimated from the ratio of the doping density to the carbon atomic density $n_C=3.82\times10^{15}~\mathrm{cm}^{-2}$ in a graphene sheet, $x^{-1}=n/n_C$. The Fermi level $E_F$ and effective mass $m_*$ are determined from the experimental density $n$ and a 3NNTB band structure calculation. The effective mass $m_*$ diverges at the Lifshitz transition.
}
			\label{tab:table1}    	
			\begin{tabular}{l*{8}{c}}
				\toprule
				& & $(e R_H)^{-1}$ & $n$ & $\mu$ & CsC$_\mathrm{x}$ & $E_f$ & $m_\ast/m_0$  \\
				& & ($\mathrm{cm}^{-2}$) & ($\mathrm{cm}^{-2}$) & ($\mathrm{cm}^{2} \mathrm{V}^{-1} \mathrm{s}^{-1}$) & & (eV) &  \\
				\midrule
				Sample 1 && $-1.4 \times 10^{14}$ & $1.4 \times 10^{14}$ & 319 & CsC\textsubscript{27.3} & 1.25 & 0.290   \\ 
				Sample 2 && $-2.0 \times 10^{14}$ & $2.0 \times 10^{14}$ & 298 & CsC\textsubscript{19.1} & 1.45 & 0.371  \\ 
				Sample 3 && $-4.0 \times 10^{14}$ & $4.0 \times 10^{14}$ & 137 & CsC\textsubscript{9.5} & 1.85 & 1.036  \\ 
				Sample 4 && $-6.1 \times 10^{14}$ & $\approx4.7 \times 10^{14}$ & 38 & CsC\textsubscript{8.1} & 1.93 - $\delta$ & -  \\ 
				Sample 5 && $+2.3 \times 10^{16}$ & $\approx4.7 \times 10^{14}$ & 59 & CsC\textsubscript{8.1} & 1.93 + $\delta$ & -  \\ 				
		\bottomrule
			\end{tabular}
		\end{center}
	\end{table*}

	Quadratic magnetoresistance $R_{xx}$ versus $B$ is observed at high field and a weak-localization peak is observed at low field. Omitting the low field regions ($|B|<0.5$~T), the magnetoresistance was fit to a quadratic form $R_{xx}(B)=R_{0}\left[1+A(\mu B)^2\right]$.  Here, $A$ is a dimensionless coefficient and $R_0$ is the zero-field resistance $R_{xx}(0)$ in the absence of the quantum effects, namely weak-localization. It is not possible to unambiguously identify the origin of magnetoresistance from the observation of a quadratic dependence on $B$. The two-band conduction model \cite{ashcroft1976solid} attributes quadratic magnetoresistance to the coexistance of a uniform distribution of two carrier types. The effective medium theory of Ping et al.\cite{Ping:2014} attributes quadratic magnetoresistance to inhomogeneity in charge carrier density in graphene. If inhomogeneity in charge density leads to electron-hole puddle formation, deviation from quadratic magnetoresistance can be observed only in the limit of large magnetoresistance  \cite{cho2008charge}, beyond what is observable in heavily doped graphene at $B=7$~T. }
	\begin{figure*}[t]
		\includegraphics[width=1.0\textwidth]{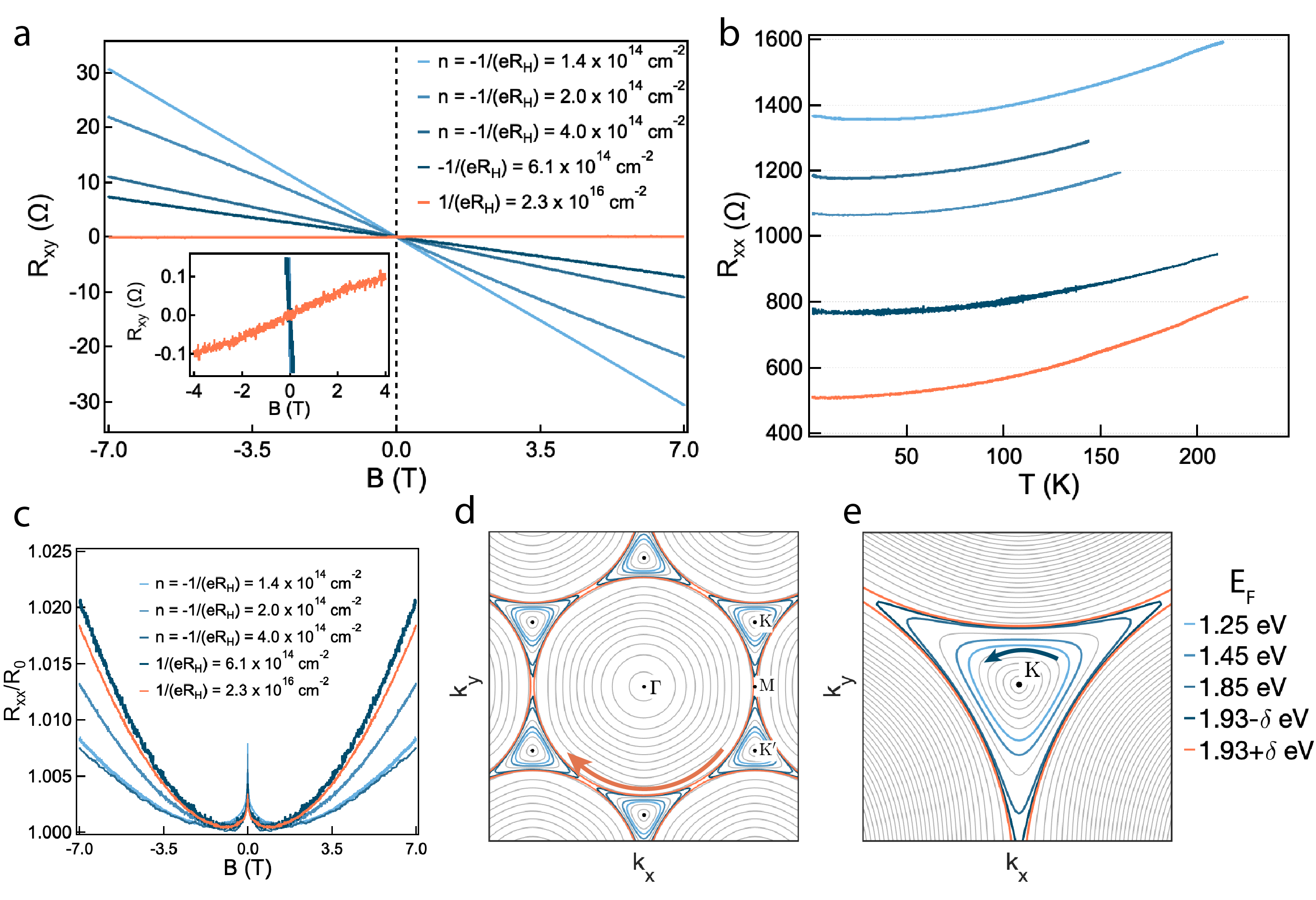}
		\caption{\textbf{High-field magnetotransport} (a) Hall resistance $R_{xy}$ versus magnetic field $B$ at $T = 1.3$~K for Cs-doped graphene samples 1-5. The inverse Hall coefficients $1/eR_H$ inferred from $R_{xy}$ are listed. Samples 1-4 show n-type $R_{xy}$, while sample 5 shows p-type $R_{xy}$, see inset. (b) The longitudinal resistance $R_{xx}$ versus temperature $T$ at $B=0$~T, showing metallic behaviour for samples 1-5. (c) Normalized longitudinal resistance $R_{xx}/R_{0}$ versus $B$ at $T=1.3$~K, with $R_0$ the zero-field resistance without weak-localization contribution. (d,e) Iso-energy contours of the conduction $\pi*$ band calculated with a third-nearest-neighbour tight binding (3NNTB) model. The Fermi energies $E_F$ and Fermi surfaces corresponding to the carrier densities $n$ of samples 1-5 are indicated. Hall coefficient and cyclotron mass inversion occurs at the Lifshitz transition, with $E_F=1.93$~eV.}
		\label{fridge}
	\end{figure*}

	\begin{figure*}[t]
		\includegraphics[width=1.0\textwidth]{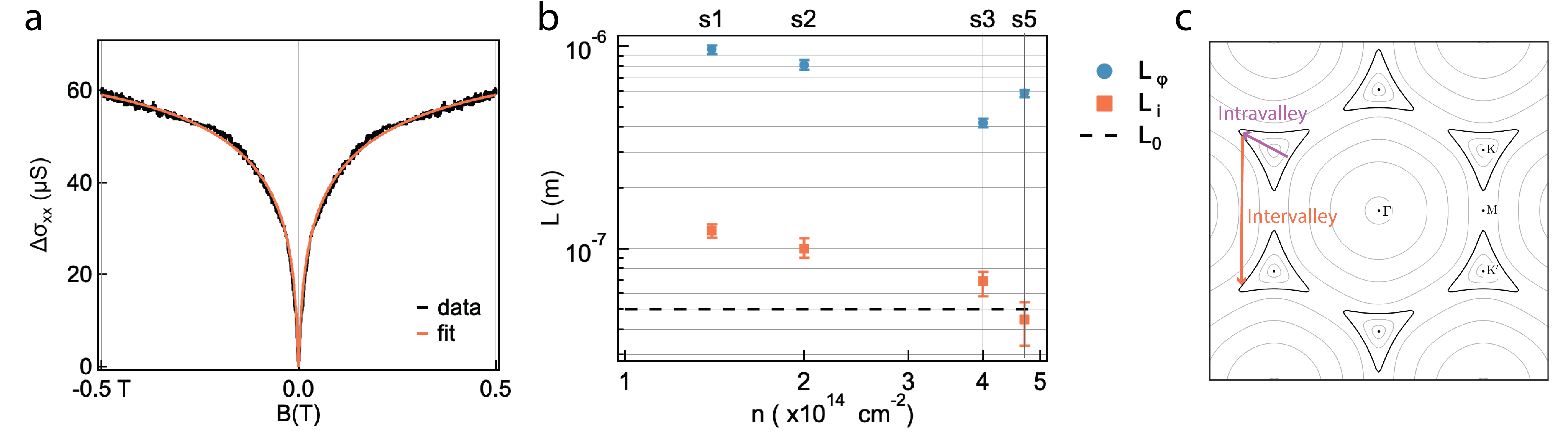}
		\caption{\textbf{Weak-localization of highly doped graphene} (a) Conductivity difference $\Delta \sigma_{xx}$ vs. $B$ of sample 5, doped above the Lifshitz transition. Experimental measurements (points) and fit to Eq. \ref{eq:WL} (solid line) are both shown. (c) Scattering lengths $L_{\phi}$ and $L_{i}$ versus density $n$ for samples 1-3, 5 at temperature $T=1.3$~K. The estimated mean-free path, $L_0$, is also indicated. (c) Schematic of intervalley scattering and intravalley scattering processes in momentum space, associated with $L_{i}$ and $L_{\ast}$, respectively.}
		\label{WL}
	\end{figure*}
	
	All samples exhibited a weak localization (WL) peak in $R_{xx}$ versus $B$ in the low-field regime (Fig.\ref{highdoping}c). WL occurs due to quantum coherent back-scattering of charge carriers, enhancing resistivity at $B=0$~T, with the effect disappearing by introduction of magnetic flux breaking the phase coherence of back-scattering events \cite{McCann:2006,fal2007weak,morozov2006strong,tikhonenko2008weak,tikhonenko2008effect,Ki:2008}. Weak-localization is sensitive to the contributions from multiple scattering mechanisms characterized by the following scattering lengths: phase coherence length, $L_{\phi}$, intervalley scattering length,  $L_{i}$, and intravalley scattering length $L_{\ast}$. {\color{black}The intervalley and intravalley scattering processes are illustrated schematically in Fig. \ref{fridge}c. The WL correction to longitudinal conductivity, $\Delta\sigma_{xx} = \sigma_{xx}(B)-\sigma_{xx}(0)$, has been analytically calculated for weakly doped graphene where the Dirac cone is an accurate model for dispersion\cite{McCann:2006,fal2007weak}, } 
	\begin{align}
		\Delta\sigma_{xx}(B)=& \frac{e^2}{\pi h}\left[F\left(\frac{8\pi B}{\Phi_0 L_{\phi}^{-2}}\right)-F\left(\frac{8\pi B}{\Phi_0 \left(L_{\phi}^{-2}+2L_{i}^{-2}\right)}\right) \right. \nonumber \\
		&\left.-2F\left(\frac{8\pi B}{\Phi_0 \left(L_{\phi}^{-2}+L_{i}^{-2}+L_{\ast}^{-2}\right)}\right)\right] \label{eq:WL}
	\end{align}	
	where $F(z)=\ln z+\psi\left(0.5+z^{-1}\right)$, $\psi(x)$ is the digamma function and $\Phi_0=h/e$ is the flux quantum. For magnetic field $|B|<0.5$~T, $\rho_{xy}(B) \ll \rho_{xx}(B)$ and thus $\sigma_{xx}(B) \approx \rho_{xx}^{-1}(B)$. {\color{black}A numerical fit of the experimental $\Delta \sigma_{xx}$ over $|B|<0.5$~T to Eq. \ref{eq:WL} was used to determine the scattering lengths, with a representative example shown in Fig.\ref{WL}(a) for sample 5 (see Supplementary Information for other samples). The scattering lengths $L_\phi$ and $L_i$ versus doping density $n$ at $T =1.3$~K are shown in Fig. \ref{WL}b. The magnetoresistance is less than 3\% of the magnitude of the WL resistance peak for $|B|\leq0.5$~T, and error bars were determined by the fit parameter deviation that increases the sum of squares error between fit and measurement two-fold over that at optimum fit value. At $T=1.3$~K, the phase coherence lengths are in the range $L_{\phi} =0.4-1.0~\mu\mathrm{m}$ and the intervalley scattering lengths are in the range $L_{i}=45-120$~nm in our heavily doped graphene samples, in general agreement with previous reports of WL in graphene \cite{morozov2006strong,Ki:2008,Baker:2012}. The shortening of $L_{\phi}$ and $L_{i}$ with increasing carrier density $n$ over the range $10^{11}-1.5\times10^{13}~\mathrm{cm}^{-2}$ has been previously observed in monolayer graphene \cite{Baker:2012}.

Being shorter than the estimated mean free path $L_0$, the intravalley scattering length $L_{\ast}$ could not be reliably determined by fit of Eq. \ref{eq:WL} to the experimental $\Delta \sigma_{xx}$ \cite{tikhonenko2008effect,lara2011disordered}. As the effective mass diverges at the Lifshitz transition, the more weakly doped samples 1-3 were used to estimate the mean free path $L_0=v_F \tau \approx 50$~nm, where $v_F = \sqrt{E_F/m^*}$ is the Fermi velocity and the scattering time $\tau$ is estimated from $\mu = e \tau / m^{*}$. Nonetheless, we may observe that trigonal warping is the dominant source of intravalley scattering at densities of $n \approx 10^{14}~\mathrm{cm}^{-2}$ with an estimated scattering length \cite{McCann:2006,Baker:2012},
	\begin{gather}
		L_{\ast}\approx L_{w}=\left(\frac{3^3}{2}\right)^{1/2} a_0 \left( \frac{\gamma_0}{E_F} \right)^2,
	\end{gather}
	where $a_0=2.46~\mathrm{\AA}$ is the graphene lattice constant and $\gamma_0 = -3.441$~eV is the nearest-neighbour hopping energy (see Supplementary Information), corresponding to $L_{w}=3-7$~nm. This model is based on a 1NN approximation that is appropriate to weakly-doped graphene where a Dirac cone approximation to energy-momentum dispersion is accurate. A theoretical treatment of the scattering rates and lengths near the vHS of monolayer graphene has yet to be developed for doping levels close to the Lifshitz transition.}

	\begin{figure*}[!]
		\includegraphics[width=1.0\textwidth]{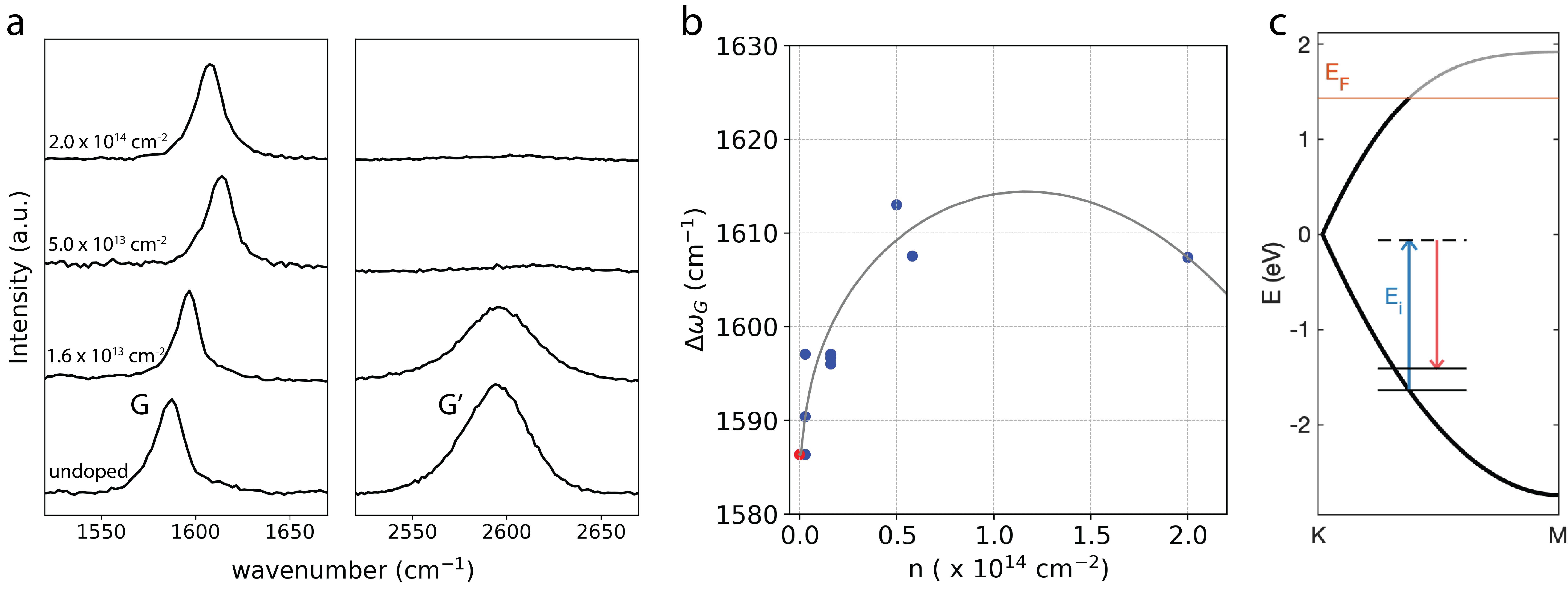}
		\caption{\textbf{Raman spectroscopy of highly doped graphene} (a) Raman spectra of doped graphene samples with varying carrier densities in the spectral regions of the $G$-peak and $G'$($2D$)-peak regions. (b) Evolution of the experimentally measured $G$-peak position $\omega_{G}$ vs. carrier density $n$ (dots, blue for Cs doped samples and red for native doping). The theoretical model of $G$-peak shift $\Delta \omega_{G}$ given by Eq. \ref{eq:Gshift} (grey line) with $\alpha = 0.36$ is in good agreement with experiment  (c) Illustration of the non-resonant Raman scattering process, with Fermi level $E_F$ corresponding to a carrier density $n=2.0 \times 10^{14}$ $\mathrm{cm}^{-2}$ (orange line), and the corresponding occupied electronic states highlighted (black) with a 3NNTB model of electronic structure. The non-resonant Raman scattering process with an incident photon energy $E_{i}=1.58$~eV is illustrated.}
		\label{raman}
	\end{figure*}

	Beyond charge transport measurements, Raman spectroscopy provides an independent experimental probe of charge carrier density via the {\color{black}measured shift of the vibrational frequencies to the effects of lattice expansion and electron-phonon coupling.} Raman Stokes spectra were measured in an series of Cs-doped graphene samples independent of those used in charge transport experiments. Fig. \ref{raman}a presents the room temperature Raman spectra of several samples with different doping densities $n$ determined from AC Hall measurements. The Raman $G$-peak position, initially at 1586~cm$^{-1}$ for pristine samples with native doping, shifts up with doping and reaches a peak value of 1613~cm$^{-1}$, followed by a slight downshift with further doping (Fig. \ref{raman}b) in qualitative agreement with previous experimental observations of ionic liquid gated graphene\cite{Pisana:2007,Das:2008,Froehlicher:2015}, Rb-doped graphene on SiO$_2$ \cite{parret2013situ}, and Cs-doped graphene on Ir \cite{hell2018resonance}, and first-principles calculations of the $E_{2g}$ $\Gamma$ phonon hardening / softening \cite{Lazzeri:2006}. Further insight can be obtained by comparison of the observed $G$-peak shift, $\Delta\omega_{G}$, relative to pristine samples with the model of Hell et al.\cite{hell2018resonance},  
	\begin{gather}
		\Delta\omega_{G}=\alpha\Delta \omega_{S}+\Delta \omega_{D}, \label{eq:Gshift}
	\end{gather}
	where the static contribution, $\Delta \omega_{S}$, is associated with lattice expansion and adiabatic electron-phonon coupling, $\alpha$ is the ratio of static $G$-peak shift of graphene in the presence of graphene-substrate interactions to that of an ideally decoupled graphene, and the dynamic contribution, $\omega_{D}$, is associated with non-adiabatic electron-phonon coupling effects. The static contribution $\Delta \omega_{S}$ is estimated with an analytic approximation to numerical calculation \cite{Lazzeri:2006}. The dynamic contribution $\Delta \omega_{D}$ is determined by numerical calculation of the electron-phonon coupling \cite{Lazzeri:2006}, using 3NNTB electronic structure \cite{gruneis2008tight} and the deformation potential, $D^2 = 63.1~\mathrm{eV^2/\AA^2}$ derived from ARPES analysis of Cs doped graphene \cite{hell2018resonance} (see Supplementary Information). A fit of the experimentally observed $\Delta \omega_{G}$ to the model of Eq. \ref{eq:Gshift}, shown in Fig. \ref{raman}b gives agreement with $\alpha=0.36$. In comparison with Cs doped graphene on Ir (111), where $\alpha=0.18$ was found \cite{hell2018resonance}, the interaction of graphene transferred to quartz suppresses lattice expansion to a lesser degree than epitaxial graphene on Ir (111). 

	Increased Cs-doping also suppresses the $G'$($2D$) peak intensity relative to the $G$ peak intensity (Fig. \ref{eq:Gshift}a), in agreement with experiments with ionic liquid gated doped graphene \cite{Das:2008,Chen:2011} and Rb-doped graphene \cite{parret2013situ}. We used pump excitation at $\lambda = 785$~nm ($E_{i}=\hbar \omega_{i} = 1.580$~eV) to minimize fluorescence. At this excitation energy, $E_{i} < 2|E_F|$ as illustrated in Fig. \ref{eq:Gshift}c, and Raman scattering is strongly detuned from resonance. Under the highly detuned conditions of our experiments, the $G'$($2D$) peak intensity is expected to be diminished to a greater extent than the $G$ peak intensity as detuning increases with doping, since the $G'$($2D$) and $G$ processes are 2nd order and 1st order, respectively. In summary, Raman spectroscopy provides an independent confirmation of the high doping achieved in Cs doped graphene produced in this study.

	\section{Conclusion}

	{\color{black}We have demonstrated a comparatively facile method utilizing an inert glove box environment for doping graphene to a high charge carrier density, up to $4.7\times10^{14}~\mathrm{cm}^{-2}$, approaching the stoichiometric limit CsC$_8$, and sufficient to observe effective mass inversion at the Lifshitz transition of monolayer graphene.} The method permits operando charge transport characterization during alkali doping, and enables further experimental characterization by methods such as magnetotransport and Raman spectroscopy. Surprisingly, this method shows that the charge densities that can be achieved by alkali doping in a diffusive vapour transport regime are similar to those achieved in UHV conditions. The charge carrier density reached by this method exceeds previous reports, and is thus anticipated to enable studies of highly doped few-layer materials, complementing ARPES studies of highly doped systems on metallic substrates. The extension of the doping method presented here to other alkali dopants and target materials and devices may provide a new means of exploring the physics of highly doped low-dimensional systems such as heavily doped black phosphorus which have been shown to exhibit bandgap modulation, closure, and inversion owing to an unusually strong Stark effect \cite{kim2015observation,ehlen2018direct}, or alkali doped C\textsubscript{60} to synthesize high critical temperature superconductors \cite{Zong:2022}. In terms of commercial applications, the flip-chip method can be implemented to dope microfabricated ion traps designed for quantum computing \cite{Auchter:2022}.

	\section{Acknowledgements}

	A.M.A., O.D., S.N.R.B. and T.S. acknowledge financial support from the Natural Sciences and Engineering Research Council of Canada (RGPIN-2018-04851,ALLRP-571923-22), Fonds de recherche du Québec – Nature et technologies (299633), and technical assistance from the Laboratoire de Microfabrication, \'Ecole Polytechnique, Montr\'eal. A.G. acknowledges Deutsche Forschungsgemeinschaft projects SE2575/4 and GR3708/4-1.

	\bibliography{doping.bib}
\end{document}